\documentclass[conference]{IEEEtran}
\usepackage{graphicx}
\usepackage{pdfpages}
\usepackage{fixltx2e}
\usepackage{amsmath}
\usepackage{epsfig}
\usepackage{tabularx}
\usepackage{url}
\usepackage{algorithm}
\usepackage{algorithmic}
\usepackage{ifthen}
\usepackage{caption}
\usepackage{subcaption}
\usepackage{subfig}
\begin{document} 
\title{AMPF: Application-aware Multipath Packet Forwarding using Machine Learning and SDN}
\author{Thomas Valerrian Pasca S, Siva Sairam Prasad, Kotaro Kataoka\\
Department of Computer Science and Engineering\\
Indian Institute of Technology Hyderabad, India\\
Email: \{cs13p1002, cs14resch01003, kotaro\}@iith.ac.in }
\maketitle

\begin{abstract}
%\boldmath
This paper proposes an application-aware multipath packet forwarding framework that integrates Machine Learning Techniques (MLT) and Software Defined Networks (SDN). As the Internet provides a variety of services and their performance requirement has become heterogeneous, it is common to come across the scenario of multiple flows competing for a constrained resource such as bandwidth, less jitter or low latency path. Such factors are application specific requirement that is beyond the knowledge of a simple combination of protocol type and port number. Better overall performance could be achieved if the network is able to prioritize the flows and assign resources based on their application specific requirement. Our system prioritizes each of the flows using MLT and routes it through a path according to the flow priority and network state using SDN. The proof of concept implementation has been done on OpenvSwitch and evaluation results involving a large number of flows exhibited a significant improvement over the traditional network setup. We also report that the port number and protocol are not contributing to determine the application in the decision-making process of Machine Learning (ML).
\end{abstract}

%\category{C.2.1}{COMPUTER-COMMUNICATION NETWORKS}{Network Architecture and Design}
%%\vspace{-0.3cm}
%\terms{Design, Experimentation, Performance}
%%\vspace{-0.3cm}
%\keywords {SDN, QoS, Application-aware, Machine Learning, Multipath Packet Forwarding  }
%%\vspace{-0.3cm}
\section{Introduction}
Internet applications and services have their own specific requirements that vary on multiple properties such as BandWidth (BW), jitter, delay, and priority. Such heterogeneity provides an opportunity to achieve better overall user experience by applying appropriate Quality of Service (QoS) in the network. Availability of multiple paths between end nodes is widely considered and implemented in modern networks, where each path may differ from other path in terms of available bandwidth and inherent delay. 

If a network is aware of state of possible end to end paths, such state can be used to intelligently assign a suitable path to each flow depending upon the requirement of the flow. However the major challenges we observe are 1) awareness of application is not easy to achieve because many applications can be integrated with a particular protocol, like HTTP meaning TCP port 80, and 2) state of network changes dynamically and static path assignment may not perform in the desired manner. In the case of conventional network, since there is no centralized mechanism which has full knowledge of the network, it is difficult to do such intelligent routing. So, here we use SDN~\cite{mckeown2009software} to choose the path intelligently and optimizing the performance of applications depending on their requirements.

In an SDN there is a centralized node which has full knowledge of the network, called as SDN controller. This SDN controller uses its knowledge to route a traffic flow. The control plane of each switch in SDN is shifted to the controller so that a centralized intelligent decision can be taken by the controller. The normal SDN also behaves like conventional network as it allocates the same path for a pair of source and destination. If a network has both application awareness as well as multipath packet forwarding then these two features can be used to classify different applications and assigning them different paths depending on their dimensions. This will increase QoS to the end-user.

In this paper, we propose a mechanism for multipath packet forwarding based on awareness of application and path state. The proposed system uses ML methods to evaluate the characteristics of flows, based on 6 out of 44 netmate flow parameters ~\cite{netmate}. Priority classes are assigned to flows depending on the flow characteristics. Now, the controller evaluates characteristics of each possible path based on parameters such as available bandwidth and delay. After getting information of all priority classes (flows are classified among these classes) and all paths, the controller assigns paths to flows based on the priority of class they belong to.

The contribution of this paper is to achieve the awareness of application in the SDN by using Machine learning technique without using Deep Packet Inspection (DPI) and integrate such a feature to QoS in a multipath network as a running system. The proposed system is expected to bring benefit to a large-scale multipath environment with QoS demands such as data center networks, telecom/data networks, and campus networks.
%\vspace{-0.3cm}
\section{Related Work}
Weiyang Mo et al.~\cite{mo2013situation} proposed a strategy to overcome traffic congestion and physical impairment in OpenFlow network. They look for an alternate path to avoid congestion by changing the path of high-data-rate IP traffic to circuit switching and also physical impairments like fiber cut. These fiber cuts can be determined by monitoring the abnormality in the path between optical nodes. But our approach proactively finds a better path and inserts the flow rules and hence avoids congestion for sensitive and delay intolerant traffic. Also, the work does not use any application classifier mechanism. 

Marc Koerner et al.~\cite{koerner2014evaluating} took a sample data-center network architecture (fat tree topology) and have proposed multiple paths between a pair of switches in inter-rack communication using a specific algorithm in the fat tree. But their multipath packet forwarding is random and does not consider QoS. Qazi et al.~\cite{qazi2013application} proposed the application awareness of the traffic. From evaluations, we find that IP address and port-based classification is not sufficient to classify the applications. 

%\vspace{-0.3cm}
\section{Application-aware multipath\\ packet forwarding (AMPF)}

AMPF introduces machine learning to classify the traffic and assigns an appropriate path to each of the flow based on their performance requirement. The idea is to provide constrained resources of a network like bandwidth and low latency paths according to their priority and class. Figure.~\ref{fig:architeture} shows the system diagram of the proposed system. A Machine Learning Trainer (MLT) and Machine Learning Classifier (MLC) are integrated into SDN controller. The MLT is used to train the classifier. The trained MLC classifies the packet into one of the predefined set of classes. Based on the class identified and policy (set by the administrator), the controller decides the path to be used for the specific application.

\subsection{ML for getting aware of applications in a flow}
MLT uses supervised learning algorithm to build the C4.5 Decision Tree (DT) classifier. A trainer is given training dataset which has the fields as given in~\cite{netmate}. The training dataset contains feature vector and corresponding class labels. MLT creates a classifier model based on training data set. MLC is generated by MLT. MLC gets a set of feature vectors comprising all the above-mentioned fields. MLC predicts the class to which the given set of features might belong. The accuracy of DT is comparatively better than that of the set of other classifiers mentioned in section 4.2.

\subsection{AMPF Controller}
\begin{figure}[!htb]
%\vspace{-0.6cm}
\centering
\epsfig{width=6cm, figure=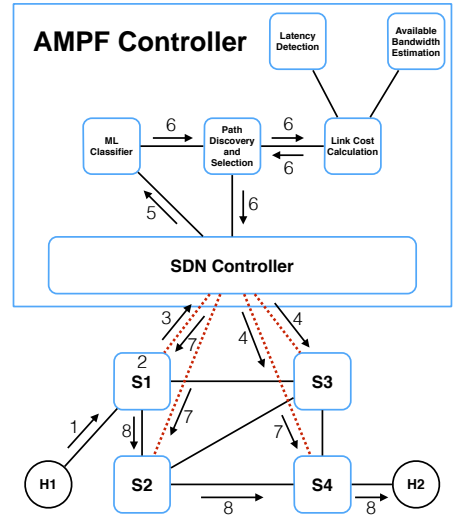}
% \includegraphics[width=0.4\textwidth,natwidth=1,natheight=1]{SystemDesign.pdf}
%  \includegraphics[scale=0.2, natwidth=1,natheight=1]{SystemDesign.pdf}
%\vspace{-1cm}
\caption{ Application-aware Multipath Packet forwarding Computation Module and its Components}
\label{fig:IMPC}
\end{figure}

AMPF controller module is a sub module in SDN controller. AMPF controller contains Latency Detection (LD) module, Available Bandwidth Estimation (ABE) module, Link Cost Calculation (LCC) module and Path Discovery and Selection (PDS) module. LD module computes the latency of all the links between switches connected to the controller. ABE module finds available bandwidth of all the links and updates available bandwidth for every new flow which is added to network. Once the latency and available bandwidth are computed link cost module computes the cost of all the links. 
AMPF controller has the complete topology of the network and it interacts with MLC and MLT. Initially, AMPF controller forwards the training dataset to the MLT for training. If training is done, then for every new flow, AMPF controller captures the feature vector of the flow and forwards the flow features to MLC to get the corresponding class label. Based on class reply from MLC, AMPF controller computes best $K$ available paths from source to destination and chooses one appropriate path for the flow. Choosing a flow is based on cost of the path and class of the application which is sent through that path. Once the path is chosen AMPF controller inserts the corresponding flow rules in all the switches which lie in that path. (From now controller and AMPF controller will be used synonymously). Figure.~\ref{fig:IMPC} shows the working of AMPF controller module.
%\vspace{-0.3cm}

\subsection{Flow of Packet in the Network} 
Figure.~\ref{fig:IMPC} shows the flow of a packet in the network. The entire network is assumed to have SDN switches and all switches are connected to one centralized controller. The flow of a packet in network is described as follows: 
\begin{enumerate}
\item A packet is sent by a host to the SDN switch.
\item SDN switch on receiving the packet checks in the flow table for a matching flow rule. If the flow rule exists then switch forwards the packet according to the flow rule.
\item If there is no flow rule in the flow table of SDN switch then the switch forwards the packet to AMPF controller.
\item AMPF controller on receiving the packet from switch computes an average length path for the new flow and writes the flow rule on all switches on the path except the source switch. These flow rules are written with HARD TIMEOUT of '$t$' seconds. Since source switch doesn't have any flow rule, it keeps on sending packet to controller. AMPF controller collects the feature vector from the packet and sends PACKET\_OUT to source switch to forward to next switch on the line. Once the AMPF controller gets first N packet of the same flow it sends all the collected feature vectors to MLC.
\item MLC on receiving the feature vector from the AMPF controller classifies it into one of the predefined classes.
\item AMPF controller receives the class label from MLC and queries the PDS module for the best path, for given priority of obtained class. It sends flow rules (FLOW\_MOD message) to all switches in the selected path. The priority of newly inserted flow is higher, so the path switching from average path in which packet was traveling to best path which is assigned will not introduce any packet loss. Assume, if a packet is in midway between source and destination and a new flow rule is written. Then, if the switch which is forwarding the packet got a new flow rule will forward in the new path. If the switch did not get any flow rule it will forward in existing path.
\item SDN switch on receiving the PACKET\_OUT message forwards the packet to the respective port. If it receives the FLOW\_MOD message it updates the flow in flow tables. 
\item SDN switch forwards the packet to the destination host.
\item Controller at $t-10$ epoch time for a flow checks the throughput experienced by a flow, by sending statistics request to switch. If bandwidth requirement is achieved, then AMPF controller will maintain the current path. Otherwise, the operation in step 6 will be conducted.
\item AMPF controller install the same rule in existing path for the flow with same priority, thereby extending the flow timeout by another \textit{t} seconds.

\end{enumerate}

%\vspace{-0.5cm}
\section{Implementation}

\subsection{Flow Rule Computation}
Algorithm 1 describes the multipath packet forwarding in PDS module, that collects data from different module and takes an appropriate decision. Initially, the AMPF controller calculates the network topology and sends LLDP packet once in '$t$' seconds to collect link statistics. AMPF controller creates a Cost\_Map and updates cost for every link using LCC module. For every new flow that arrives at AMPF controller, it collects the feature vector and gets it classified as mentioned in section 4.2. CLASS (CLS) is the variable storing the class value. Yen $K$ Shortest Path (YKSP) algorithm~\cite{yen1971finding} is used to find $K$ paths from source to destination of a flow. The bandwidth requirement of the path can be computed from the feature vectors. AMPF controller iterates through all the available paths to find the set of feasible path that can carry the flow. Once the set of feasible paths is found, Class\_interval (CI) is calculated as a fair division of feasible paths among all class ($N_{c}$) of flows. To find the path corresponding to $CLS*CI$ index, need to iterate through the set of all feasible paths ($N_{fp}$). By writing flow rule in that path and updating the cost of links, the path of a flow will change with information collected from the flow . This cost updating will be a good approximation but the actual cost can be obtained once an LLDP is sent out. A new flow is inserted on selected path. The flow rule inserted must have higher priority so that it will replace the flow with lower priority inserted previously. 
\label{sec:algo-summary}
\begin{algorithm}[htb!]
\begin{algorithmic}[1] 
\STATE { Controller gets the topology from Topo Manager and computes link cost by sending LLDP.}
\STATE{Every new flow, PACKET\_OUT first 50 packets on average path. Compute flow features and sends to MLC. MLC gives predicted class CLS based on feature vector}
\STATE{PDS finds K shortest path for given source to destination (YKSP), Find feasible path among them with bandwidth and delay as constraints}
\STATE {$CI = \frac{N_{fp}}{N_{c}} $}
\IF{ $ CLS*CI \leq index\, of\, feasible\, path $}
\STATE{Assign flow to the path, update link cost and break}
\ENDIF
\STATE{ Write corresponding flow rules on all the switches in the path}
\STATE {Check Epoch time, and if the flow met the bandwidth delay guarantees, maintain the same path, or reroute based on previous steps}
\STATE{ end}
\end{algorithmic}
\caption{Algorithm to forward a flow \label{algo: Algorithm to forward a flow }}
\end{algorithm}
% \subsection{Parameters to Classify a Flow}
\subsection{Application Classification}
\begin{figure}[htb!]
\centering
\epsfig{width=7cm,  figure=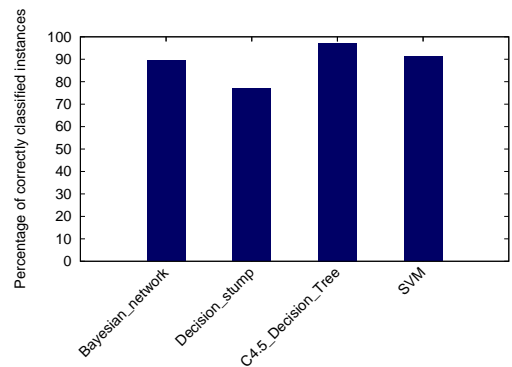}
%\vspace{-0.6cm}
\caption{Comparision of MLT}
\label{fig:ML_technique}
\end{figure}

MLC module implements the DT MLC. In~\cite{kim2008internet} it has been shown that Support Vector Machine (SVM) is the best classifier for classifying the network traffic (flows). By using correlation-based filters the number of parameters to be considered for classification is reduced in between 6 and 10 from 37. But the authors have taken protocol, ports, TCP flags and packet size information for creating training data set. Authors~\cite{kim2008internet} of  have considered port number also for classification and presented that SVM would result in poor classification if port number is not used. In TCP with Secure Sockets Layer (SSL), the destination port number is same and it doesn’t contribute to classification significantly, hence we focused on using other metrics such as Interpacket interval, Packet length. In~\cite{kim2008internet} the performance analysis of machine learning algorithm has been conducted on Naive Bayes, Naive Bayes Kernel Estimation, C4.5 DT, Bayesian Network, SVM, K-Nearest Neighbors and Neural Network. Among which SVM has achieved the highest overall accuracy while classifying. SVM classifier achieves more than 98.0\% average accuracy on all traces with 5,000 training flows, which amount to only 2.5\% of the size of the testing sets. But without port number their classification accuracy went to 56 - 70\%. Our Figure ~\ref{fig:ML_technique} based on flow capture of YouTube, Facebook, Skype, Dropbox and Copy reveals that C4.5 DT gives the highest classification accuracy compared to the other algorithms.

%\vspace{-0.4cm}
\subsection{Multipath Packet Forwarding}
%\vspace{-0.3cm}
\subsubsection{Measuring Latency}
We use the paper~\cite{phemius2013monitoring} for calculating one-way latency of a path and use LLDP packet instead of using broadcast packet, LLDP is much efficient to measure latency with less packets. LD module computes the latency of all the links in the network. To compute latency of any link in the SDN based network we have used the solution proposed by~\cite{phemius2013monitoring}. The controller sends reference packet with broadcast destination as MAC address. The packet payload contains port number and time stamp of packet creation. The packet is sent to a switch with a PACKET\_OUT message instructing the switch to forward the packet to a particular output port. When the packet reaches the next switch, since the switch does not have corresponding flow entry, it sends the packet back to controller. On receiving the packet with source MAC address as reference, controller finds the Entire Trip Time (ETT) $T_{total}$ by subtracting the packet creation time from reaching time. ETT (Controller - Switch 1 - Switch 2 - Controller). One way ETT (Controller - Switch 1- Controller) is found by sending STATISTICAL\_ REQUEST message to switch for which switch will respond with STATISTICAL\_ 
RESPONSE message in time $T_{s1}$. Similarly from switch 2 ETT $T_{s2}$ can be found in~\cite{phemius2013monitoring}. 
%\vspace{-0.3cm}
\begin{equation}
Latency (s1,s2) = T_{total} -\Big(\frac{T_{s1}}{2}\Big) - \Big(\frac{T_{s2}}{2}\Big) 
\end{equation}

Our system implementation of LD module is slightly different. Instead of sending explicit control packets for checking the links, we are using Link Layer Discovery Protocol (LLDP) packets which are sent periodically to check the link status and also to get the link delay. This reduces the number of control packets sent into the network. 

\subsubsection{Estimating Link Cost and Assigning Path to Flow}
LCC module is responsible for finding link cost over each link. Link cost is derived from LD module. LCC module also finds bandwidth available from the pilot packets. From the pilot packet size and latency, available bandwidth is calculated. Available bandwidth is used in finding whether a flow can be accompanied in a route or not.

YKSP algorithm is used for computing multiple paths from source to destination. YKSP has computational complexity $O(Kn^{3})$. YKSP is better than Pollack's~\cite{pollack1961letter}, Bock, Kantner and Haynes'~\cite{bock1958algorithm}, Clarke, Krikorian and Rausan's~\cite{clarke1963computing} and Sakarovitch's~\cite{sakarovitch1966k}. There are many variants of YKSP algorithm. But we have chosen basic YKSP implementation. YKSP algorithm needs the network topology; all the switches and corresponding links in graph format. The link cost of every link $LC_{i}$ is given as weight of the link in graph. YKSP finds the shortest path in the graph using Dijkstra's algorithm. If shortest path is found it removes one link in shortest path and tries to find the shortest path again and so on which will give second shortest path. The algorithm iterates up to $K$ paths. Once $K$th path is found, cost for every path is computed.
The cost of a link depends on latency of the link ($\ell$) and Available Bandwidth ($AB$) of that link. The latency part says how much delay that particular link will cause and available bandwidth part says why this link should be preferred over other links. If two links have the same latency the preference of one link over the other will be solved by available bandwidth part. Available bandwidth $AB_{i}$ is normalized to have a fair estimate of link. 
Normalized Available Bandwidth for $i^{th}$ link is given by($NAB_{i}$).

%\vspace{-0.3cm}
\begin{equation} 
NAB_{i}=\frac{AB_{i}}{Max(AB)} \forall i
\end{equation}
%\vspace{-0.4cm}
\begin{equation}
LC_{i}=\lambda_{a}*\ell_{i} + \lambda_{b}*\frac{1}{NAB_{i}}
\end{equation}

Here, $\ell_{i}$ is the link latency between switch a and switch b $Latency(s_{a},s_{b})$. where $a$ and $b$ are neighboring switches.
$\lambda_{a}$ and $\lambda_{b}$ are appropriate weights given to emphasize the corresponding variable impact on cost. Cost of a path p ($CP_{p}$) is always the sum of link cost ($LC_{i}$) of all links in that particular path $\nu_{p}$.

%\vspace{-0.4cm}
\begin{equation}
CP_{p} =\sum\limits_{i}^{\nu_{p}} LC_{i} 
\end{equation}

Based on flow priority and cost of a path, a flow is sent through the path (as mentioned in Algorithm. 1).

%\subsubsection{Avoiding Broadcast Storm Intelligently}
%Complex topology (topology with loop) creates ARP broadcast storm. Broadcast storm can be resolved with well-known STP (spanning tree protocol), but Floodlight-0.90 (latest) does not have the implementation.\\
%\indent We resolved this problem in an effective way. A Map is created with [IP, TCP] tuple of source packet as key and set of switches from which controller gets ARP request as value of map. If an ARP request comes from a switch which has a corresponding key field in the map, controller checks the set of visited switches. If the switch is not in visited switches entries then it checks all the output ports of that particular switch and floods the packet only to those switches which are the next hop of current switch, but not present in visited switches. This topology is adopted instead of spanning tree because even if a link in spanning tree fails out, the entire topology is split-up such that some nodes might not be reachable with current spanning tree. In that scenario, our method will find another way to reach the destination.

\subsection{Data Structure}
Hash Map and concurrent hash maps are used throughout the implementation. Concurrent hash maps are used when two or more threads are changing the same hash map. For resolving ARP broadcast storm, set of visited switches is stored in a concurrent hash map so that any thread of controller can change it independently. Similarly, latency map which is used to store latency is also a concurrent hash map with key as link and value as latency. Once LLDP packet is received at the controller, it updates the map independent of latency map used by other modules for computing path.

\begin{table}[htb!]
%\vspace{-0.3cm}
\centering
\caption{Implementation Details}
\begin{tabular}{|l|l|} \hline
\textbf{Classification}&\textbf{Details}\\ \hline
Operating System & Mininet 2.2.0 \cite{mininet} \\ \hline
Framework for developing & \\
openflow controller & Floodlight 0.90 \cite{floodlight} \\ \hline 
Software of openflow-switch & Openflow-1.0 \cite{openflow} \\ \hline
Language of Development & JAVA\\ \hline
Machine Learning Classifier & C4.5 Decision Tree\\ \hline
\hline\end{tabular}
%\vspace{-0.6cm}
\end{table}

\section{Evaluation and Results}

\subsection{Test bed setup} 

\begin{figure}[htb!]
\centering
\epsfig{width=5cm, angle =90, figure=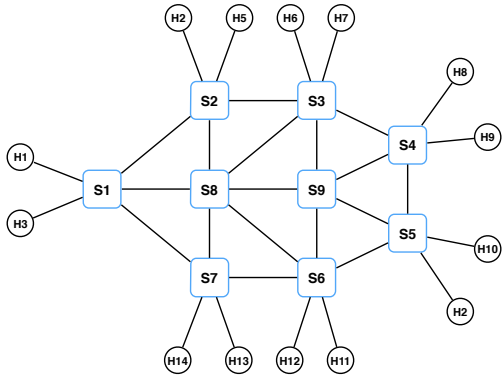}
\caption{System diagram of the proposed system}
\label{fig:architeture}
%\vspace{-0.5cm}
\end{figure}

The proposed system is evaluated over testbed described in Figure.~\ref{fig:architeture}. For the test bed, we have assumed that hosts, Host H1 and H2 are only nodes which are taking part in data transmission and reception while other nodes are idle.
The evaluation was carried out on a general Internet scenario. 
The link bandwidth between every connected switch is taken as 32 Mbps and the link bandwidth between switch and host is taken as 1 Gbps each.
This set up was evaluated on four flows namely Class 1 (Real-time Traffic), Class 2 (Buffered/file transfer), Class 3 (Web Browsing) and Class 4 (Restricted/File Transfers).
The characteristics of the flows are summarized as Tables~\ref{tab:internet} and~\ref{tab:delay}.

\begin{table}[htb!]
\centering
\caption{Testbed Flow Characteristics}
\begin{tabular}{|c|c|c|c|c|} \hline
\label{tab:internet}
\textbf{Class}&\textbf{\parbox{1.4cm}{Min\\BW}}&\textbf{\parbox{1.4cm}{Max\\BW}}&\textbf{\parbox{1.5cm}{Jitter\\Tolerance}}&\textbf{\parbox{1.5cm}{Delay\\Tolerance}}\\ [1ex]\hline
1 & 400M & N/A & Low & Low\\ \hline
2 & 200M & 400M & Mid & Mid\\ \hline
3 & 100M & 200M & Mid & Mid\\ \hline
4 & N/A & 100M & Mid & Mid\\ \hline
\hline\end{tabular}
%\vspace{-0.5cm}
\end{table}
% ..................................................
\begin{table}[htb!]

\centering
\caption{Application Delay and Bandwidth Requirements}
\begin{tabular}{|c|c|c|} \hline
\textbf {Application} & \textbf{Acceptable Delay} & \textbf{\parbox{2.1cm}{Min.BW\\Requirement}}\\ \hline
Class 1 & 20 msec & 10 Mbps\\ \hline
Class 2 & 40 msec & 5 Mbps\\ \hline
Class 3 & 60 msec & 2 Mbps\\ \hline 
Class 4 & best effort & 1 Mbps\\ \hline
\hline\end{tabular}
\label{tab:delay}
%\vspace{-0.6cm}
\end{table}

% \begin{figure}[htb!]
% \centering
% \epsfig{width=7cm, height =9cm, figure=images/lldpFreq.eps}
% \caption{Convergence of link load value for different frequency of LLDP}
% \label{fig:Monitor}
% \end{figure}

All switches are connected to AMPF controller with a dedicated link and the entire setup runs in mininet. The above table has been scaled down for evaluation purpose. Acceptable delay and minimum bandwidth requirement for the classes are shown in Table~\ref{tab:delay}. The acceptable delay is for UDP traffic whereas minimum bandwidth is the requirement of TCP traffic. In our evaluation, a Class may be either TCP or UDP.
% Minimum bandwidth requirement is been set to 10 Mbps for Class 1, 5 Mbps for Class 2, 2 Mbps for Class 3 and 1 Mbps for Class 4.
We evaluate the proposed system by observing the quality of communication using bandwidth, packet loss, jitter and delay for each class of flows. For achieving this, the network is first loaded with 8 UDP flows (as background traffic). Observable flows are started at a random interval within 30 sec. 
% The network stabilizes within 10 sec i.e., more accurate link information is obtained within 10 sec as shown in Fig.~\ref{fig:Monitor}. This delay in obtaining network information can also be further optimized; more information is discussed in section E.
All the parameters of the network (Throughput, Jitter, and latency) are tested for fixed source and destination pair. 
The flow rules expire every 100 seconds. To observe the changes in the network parameters like throughput the flows should live after the flow expiry periods. In our implementation, each flows lifetime was 1000 seconds, which provided each flow as well as the network, enough time to stabilize and analyze the behavior of that flow.

\subsection{Reducing the impact of path switching}
One of the problems faced in the evaluation setup was that of a substantial packet loss, whenever the new flow rules were installed. This was handled by queuing the packets at the switches. This drastically reduced the packet loss to a negligible amount of 0.001\%.
Every flow is checked for its minimum bandwidth requirement and rerouted if they didn't satisfy the flow requirement. The path changing happens just before flow expiry time to avoid TCP timeout for the packets queued in the buffer. Since path changing and flow rule insertion will take significant time order of few msec. This time increases the RTT and reduces the TCP widows size thereby overall throughput of that particular class comes down.

\subsection{Metrics of evaluation}
The system is evaluated in following modes.
\subsubsection{Multipath packet forwarding without application awareness}
Any new flow which arrives at the controller is sent out through one of the K available shortest paths picked at random. This choice might not give the expectation of least jitter for the sensitive application. 

\subsubsection{Multipath packet forwarding with application awareness}
Prioritizing the flow and assigning appropriate path is like the one-to-one mapping of the highest priority to the lowest latency path, the second highest priority to the second lowest latency path and so on. However, this assignment sometimes makes the application to suffers more jitter when two or more low latency paths share a set of common edges. As LCC module is updated once in \textit{t} seconds is used to compute $K$ shortest paths. Any new flow, that arrives within \textit{t} seconds takes the next shortest path. Based on the required bandwidth of the flow and classification of ML, the path is assigned after~\textit{t} seconds by observing first 50 packets of the flow.

\subsection{ Application Classifications}
\begin{table}[htb!]
%\vspace{-0.4cm}
\centering
\caption{Application Classification}
\begin{tabular}{|l|c|} \hline
\textbf {Application} & \textbf {Class}\\ \hline
Skype & 1 \\ \hline
Youtube, Google Docs & 2 \\ \hline
Gmail, Facebook & 3\\ \hline 
Dropbox, Copy, FileZilla, Torrent Client& 4\\ \hline
\hline\end{tabular}
\label{table:app_priority}
%\vspace{-0.3cm}
\end{table}
We have collected data from set of 10 applications to train the system. They are Gmail, Facebook, YouTube, Skype, Dropbox, Copy, Google Drive (docs and related stuff), Google Maps (Location update), FileZilla (File Client) and Torrent Client. The flow classification is based on $ 7 $ parameters. 
Classifier is trained with $ 500 $ flows. Classifier is also tested with $ 100 $ flows. The classifier gave more than $ 98 \%$ accurate classification. Classifier works as expected by~\cite{kim2008internet}. We have assigned priority to application as shown in Table ~\ref{table:app_priority}.

\subsection{Experiments}
%\vspace{-0.3cm }
\subsubsection{Jitter Analysis using UDP}

\begin{figure}
\begin{minipage}{1.2 in}
\epsfig{width=4.6cm,  figure=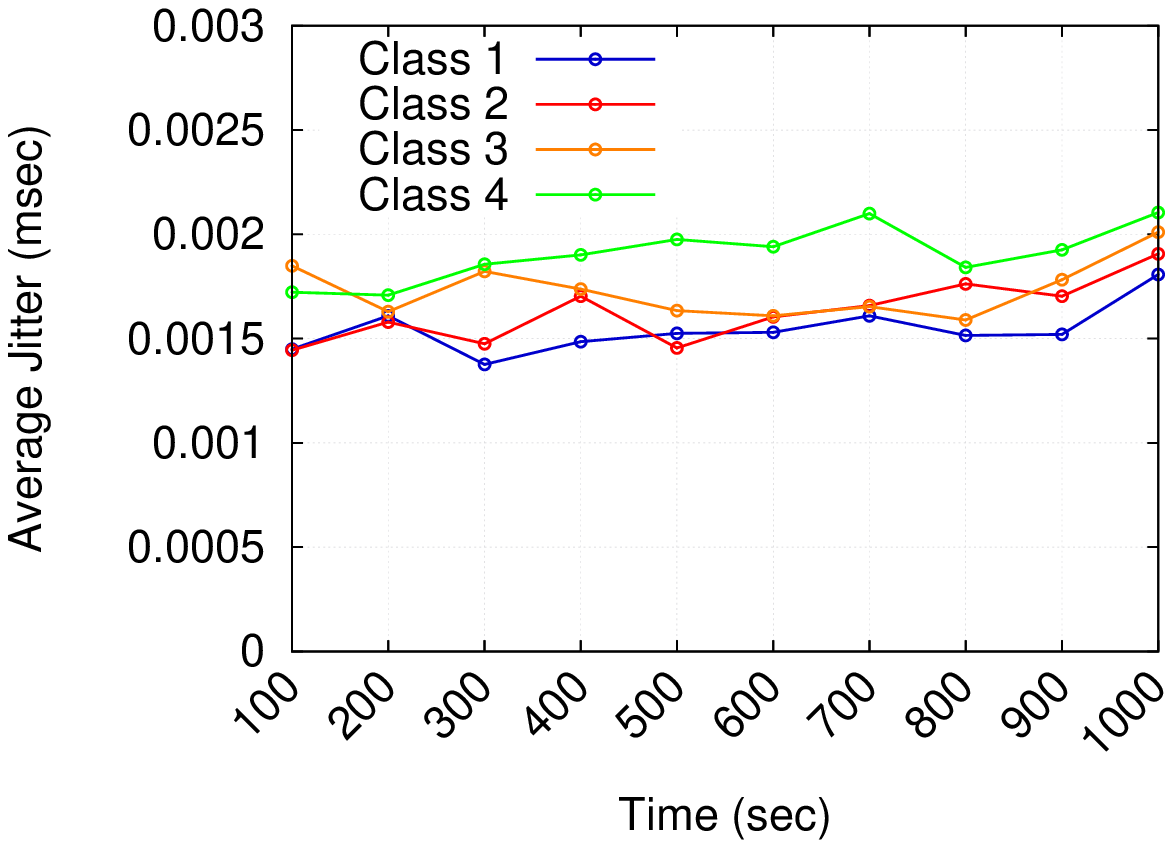}
%\vspace{-0.3cm}
\subcaption{App. Aware\\}
\label{fig:Appaware_UDP_2_jitter}
\end{minipage}
\hspace{0.5cm}
\qquad
\begin{minipage}{1.2 in}
\epsfig{width=4.6cm , figure=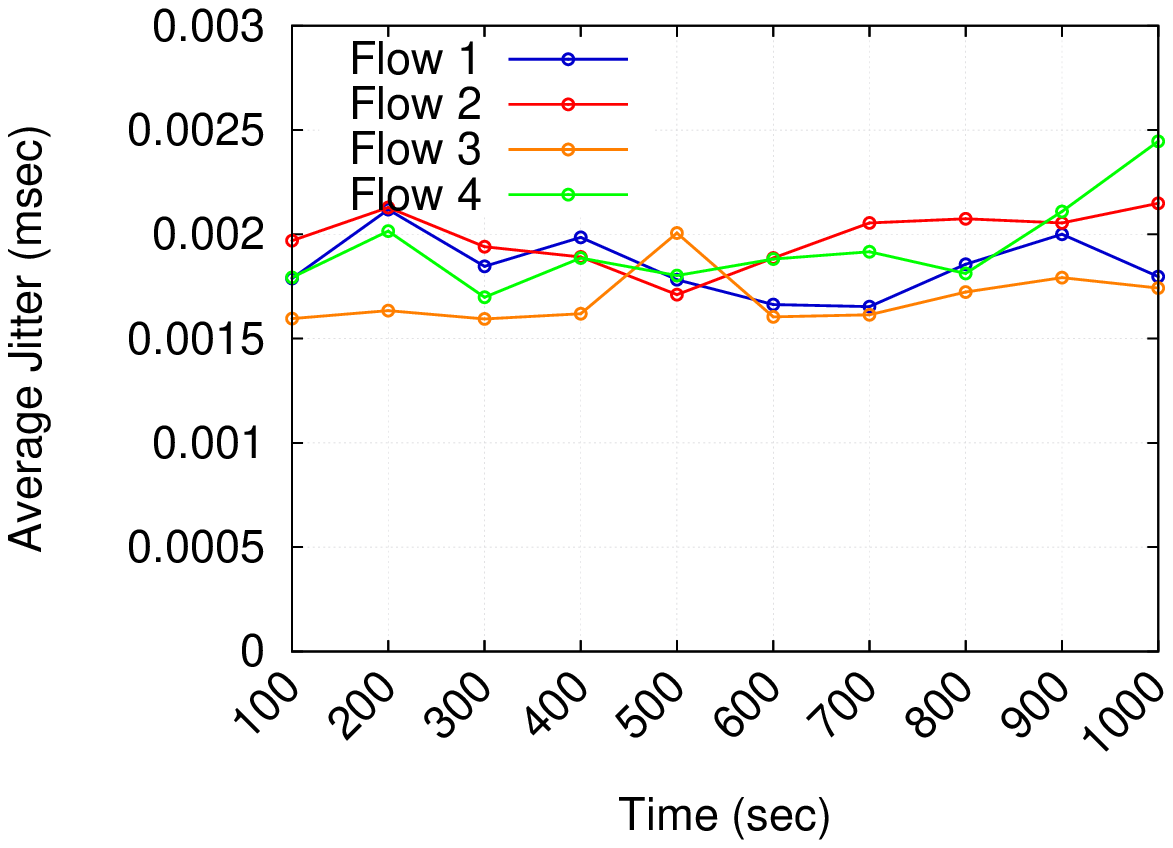}
%\vspace{-0.3cm}
\subcaption{App. Unaware\\}
\label{fig:Appunaware_UDP_2_jitter}
\end{minipage}
%\vspace{-0.2cm}
\caption{Class priority versus per class jitter, two way UDP}
%\vspace{-0.5cm}
\end{figure}

This experiment verifies that the higher priority traffic experiences comparatively lesser jitter which introduce stability of services. H1 sent out 4 UDP flows to H2 and H2 sent out 4 UDP flows to H1. Background traffic on all inter-switch links connecting to Switch 1 and Switch 5. Delay requirements (time within which the packet should be reached to destination) of the flows are shown in Table~\ref{tab:delay}. Here, we have calculated average jitter of all flows in the interval of 100 seconds. In the Figures. 4, 5, 6 and 7 App. Aware and App. Unaware means Application-aware and Application-unaware respectively. Figure. ~\ref{fig:Appaware_UDP_2_jitter} and~\ref{fig:Appunaware_UDP_2_jitter} show results of this experiment. 
The X-axis shows the time interval at which jitter is measured while Y-axis shows the average jitter measured in that time interval. In Figure.~\ref{fig:Appaware_UDP_2_jitter}, application awareness has been taken into consideration and hence, paths are assigned to flows based on the priority of the class to which the flows belong. In Figure.~\ref{fig:Appaware_UDP_2_jitter}, we can observe that high priority classes get less jitter because APMF lets them go through low latency paths. Here, class~1 (highest priority class) experiences lowest jitter while class~4 (lowest priority class) experiences the highest jitter. 

In case of Figure.~\ref{fig:Appunaware_UDP_2_jitter}, we have not considered the application awareness. Therefore, paths are assigned to flows randomly due to lack of prioritization. Since, there is no classification of flows in Figure.~\ref{fig:Appunaware_UDP_2_jitter}, therefore, we have used flow number here. But, the flow number in Figure.~\ref{fig:Appunaware_UDP_2_jitter} corresponds to class number in Figure.~\ref{fig:Appaware_UDP_2_jitter}, i.e., flow~1 belongs to class~1. Similarly, flow~2, 3 and 4 belong to class~2, 3 and 4 respectively. Figure.~\ref{fig:Appunaware_UDP_2_jitter} shows that jitter does not depend on type of flow (due to random path allocation) and thus flow~1 (class~1 flow) does not experience lowest jitter which results in QoS degradation. Since the network is lightly loaded (only UDP traffic), we can see that in both cases, average jitter lies between 0.0014ms to 0.0024ms, which is quite low. The jitter value will increase as the network load will increase. In further experiments, we have considered a heavily loaded network (both TCP and UDP traffic).

%\vspace{-0.5cm}
\subsubsection{Throughput Analysis using TCP}
\begin{figure}[htb!]
\begin{minipage}{1.2in}
\epsfig{width=4.6cm ,figure=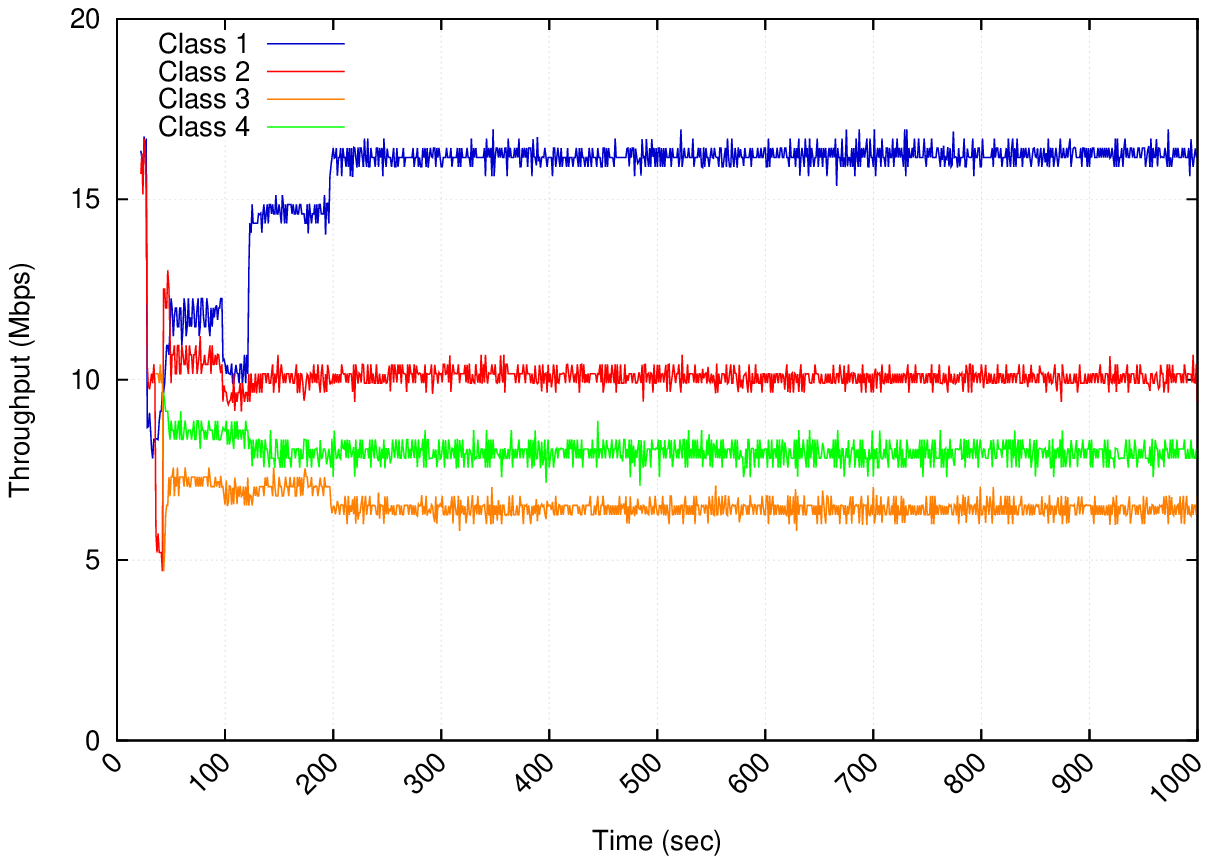}
%\vspace{-0.3cm}
\subcaption{App. Aware}
\label{fig:appaware_tcp2_throughput}
\end{minipage}
\hspace{0.5cm}
\qquad
\begin{minipage}{1.2in}
\epsfig{width=4.6cm, figure=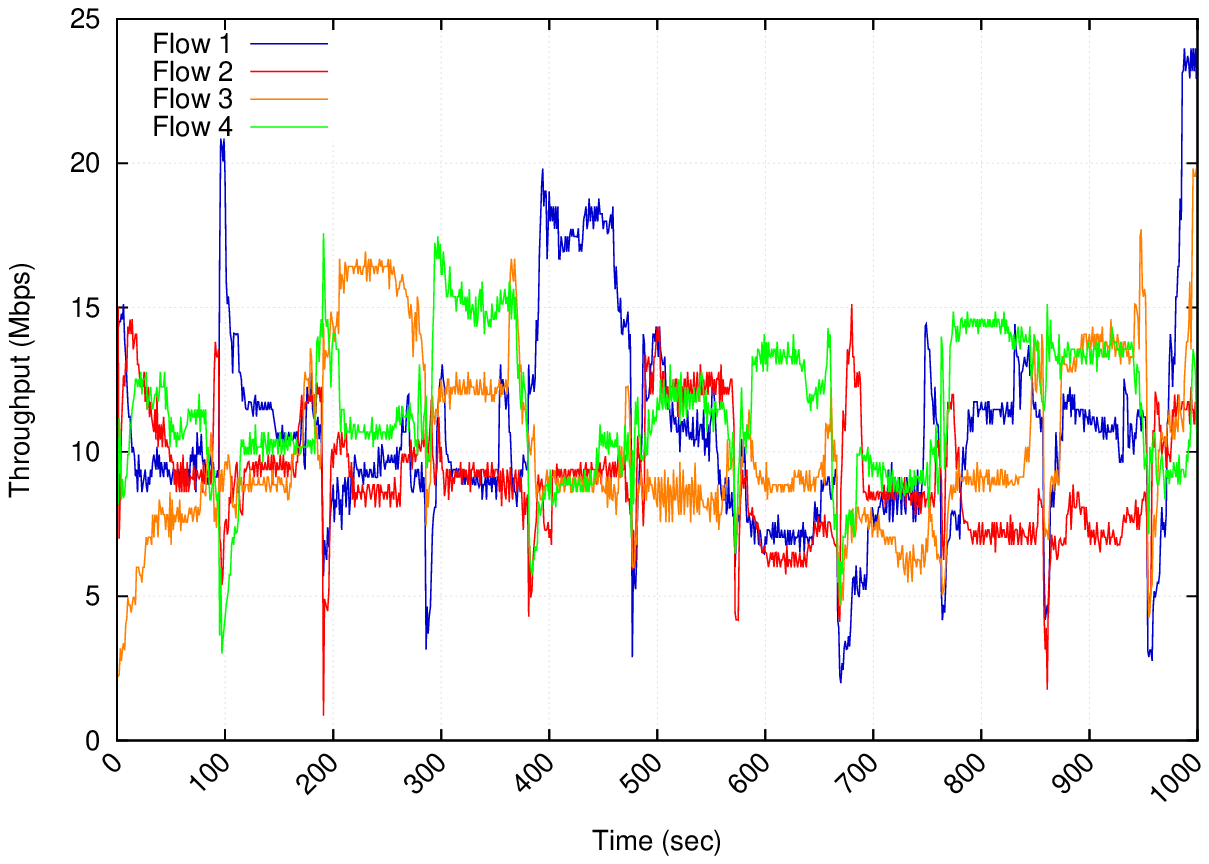}
%\vspace{-0.3cm}
\subcaption{App. Unaware}
\label{fig:appunaware_tcp2_throughput}
\end{minipage}
%\vspace{-0.2cm}
\caption { Class priority versus per class TCP Throughput, two way TCP}
%\vspace{-0.5cm}
\end{figure}

H1 sent out 4 TCP flows to H2 and H2 sent out 4 TCP flows to H1. Table~\ref{tab:delay} shows the bandwidth requirements of the flows set for experimental purpose. Each flow starts at random between \textit{t}=0 to \textit{t}=50ms with a big TCP window size and reduces according to the network. As per AMPF technique, Class~1 flows will be pushed into the path where there is least traffic so the TCP window of Class~1 grows higher whereas other flows are limited from growing further. This is because the route which other applications have chosen would have been occupied by background flows. This makes Round Trip Time (RTT) to increase for less priority applications and thereby bringing down the window size. After every 90th second, the controller checks that if the flow is not achieved the required throughput then it is rerouted by the controller. Because of this proactive routing which is done at every 90th second based on flow stats, i.e., before expiry of the flows, the new flow rules is being installed makes the flow not to suffer much because of the TCP behavior of reducing the TCP Window. This is the reason for almost no fluctuation in TCP throughput. The point to be noted that in Figure.~\ref{fig:appaware_tcp2_throughput}, Class~4 gets more throughput than Class~3 and that is because Class~3 and Class~4 are satisfying their bandwidth guarantees as we did not specify any upper limit for bandwidths. Hence, their flows are not altered. If flows did not satisfy the minimum bandwidth requirement, they would have been rerouted.

Figure.~\ref{fig:appunaware_tcp2_throughput} depicts the results of the experiment for the same configuration as above, but in a scenario in which the flows are not classified (without application awareness, as in case of Figure.~\ref{fig:Appunaware_UDP_2_jitter}). As a result, they are not routed to correct path and every flow affects the other flow in the same path. After every 100 seconds of time window, each flow expires and is placed in random path. The idea of class number and flow number in Figure.~\ref{fig:appaware_tcp2_throughput} and Figure.~\ref{fig:appunaware_tcp2_throughput} is same as Figure.~\ref{fig:Appaware_UDP_2_jitter} and~\ref{fig:Appunaware_UDP_2_jitter}.

\subsubsection{Throughput Analysis using TCP and UDP}

In this experiment, we observe the behavior of the proposed system in a more heterogeneous network.
This section of evaluation uses a similar setup as before, but this time, 4 TCP and 4 UDP flows are sent from H1 to H2 and similarly 4 TCP and 4 UDP flows are sent from H2 to H1. Here, throughput is calculated only for TCP flows.

\begin{figure}[htb!]
\begin{minipage}{1.2in}
\epsfig{width=4.6cm ,figure=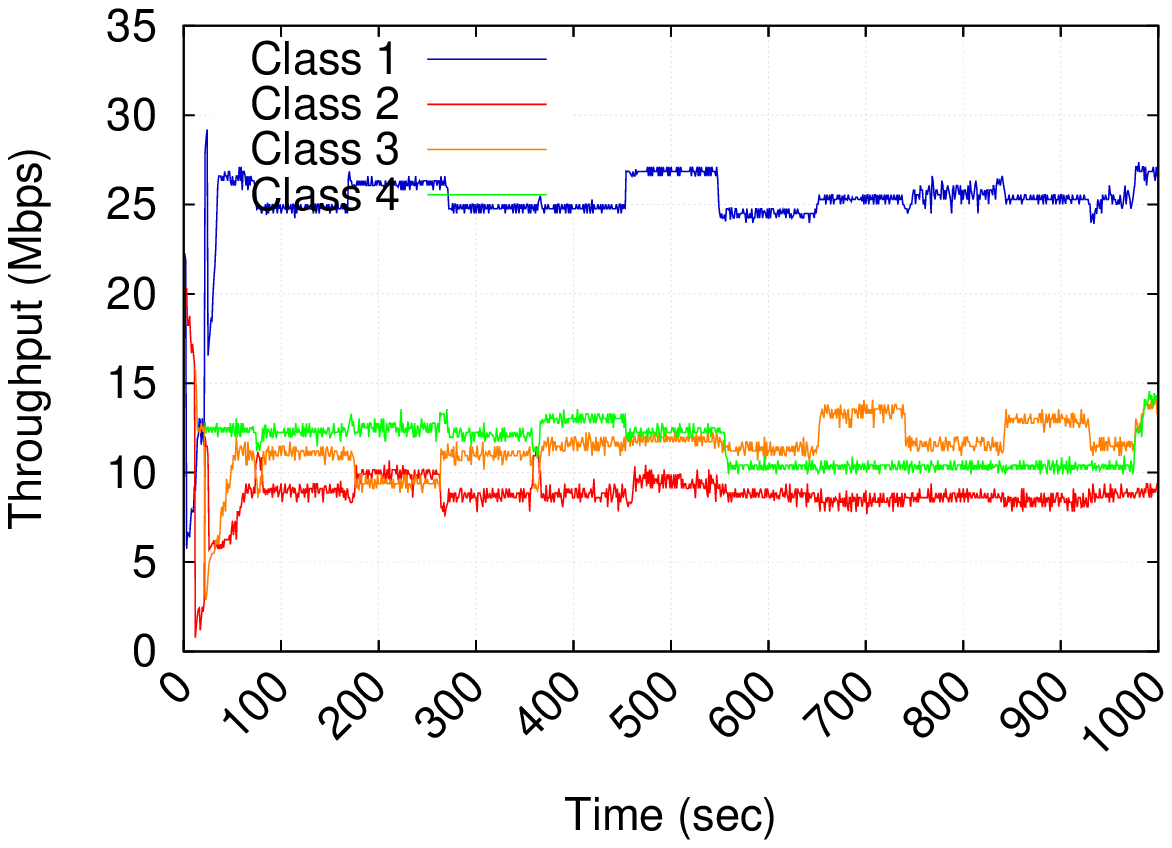}
%\vspace{-0.3cm}
\subcaption{App. Aware}
\label{fig:throughput_2_TCP_UDP}
\end{minipage}
\qquad
\hspace{0.5cm}
\begin{minipage}{1.2in}
\epsfig{width=4.6cm, figure=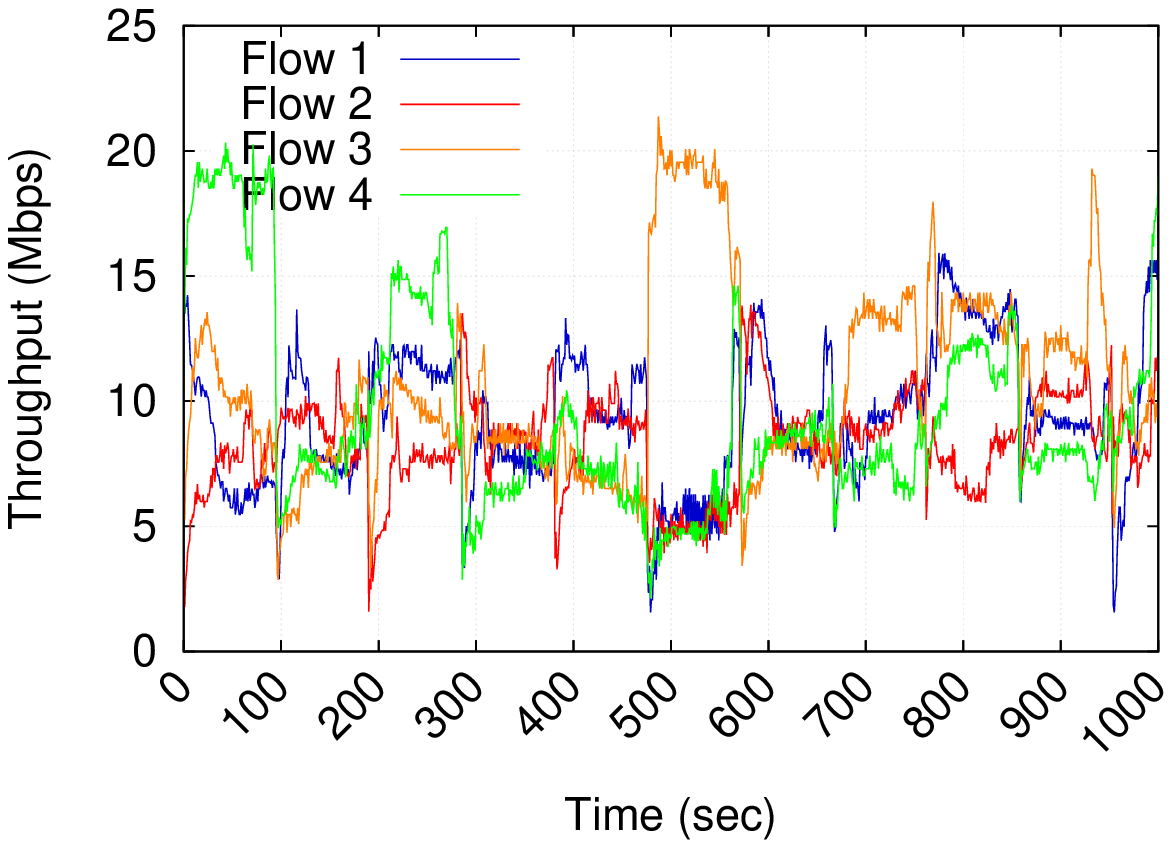}
%\vspace{-0.3cm}
\subcaption{App. Unaware}
\label{fig:unaware_throughput_2_TCP_UDP}
\end{minipage}
%\vspace{-0.2cm}
\caption {Class priority versus per class TCP Throughput, two way UDP and TCP}
%\vspace{-0.5cm}
\end{figure} 

In Figure.~\ref{fig:throughput_2_TCP_UDP}, Class~1 gives higher throughput follows the same explanation as above, rest of the classes are satisfied at much lower level hence rerouting is not done.
In Figure.~\ref{fig:unaware_throughput_2_TCP_UDP} flows are rerouted not based on class requirement and hence, they do not give a clear class wise throughput.

\subsubsection{Jitter Analysis using TCP and UDP}

\begin{figure}[htb!]

\begin{minipage}{1.2in}
\epsfig{width=4.6cm ,figure=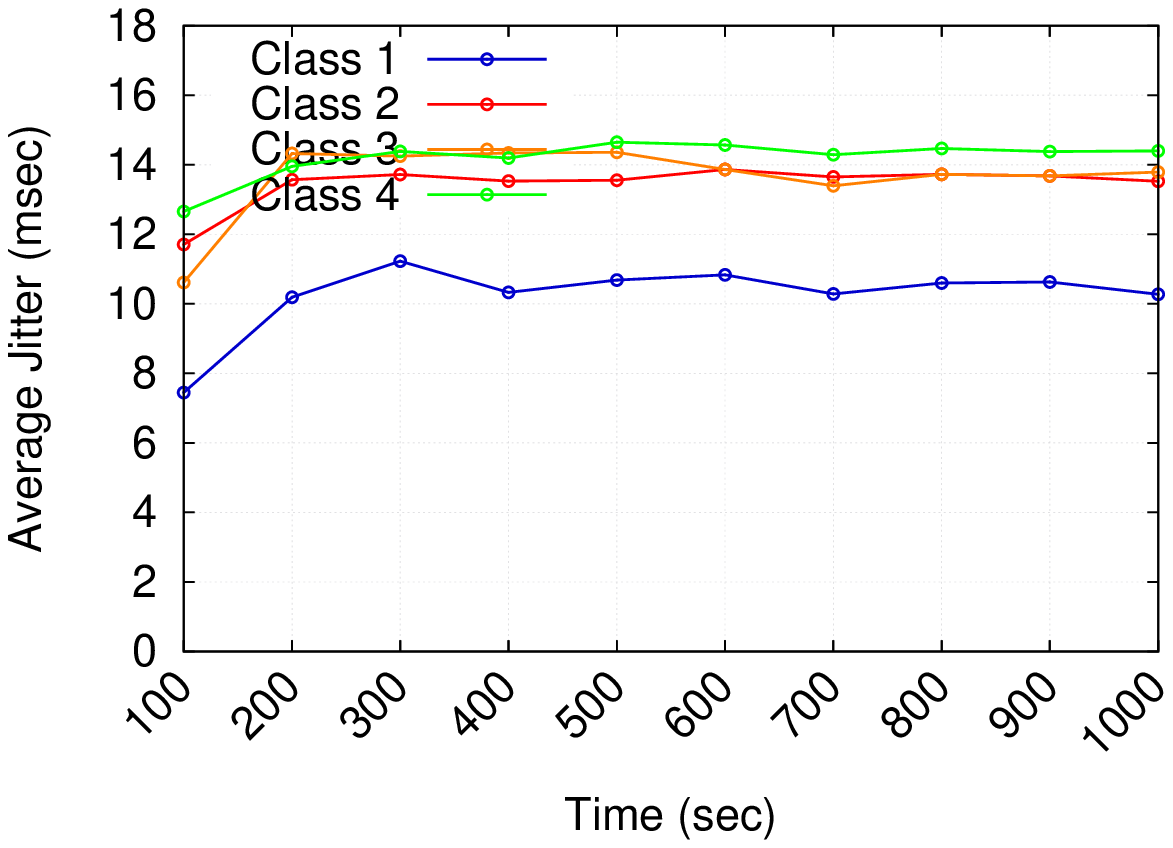}
%\vspace{-0.3cm}
\subcaption{App. Aware}
\label{fig:delay_2_TCP_UDP}
\end{minipage}
\hspace{0.5cm}
\qquad
\begin{minipage}{1.2in}
\epsfig{width=4.6cm, figure=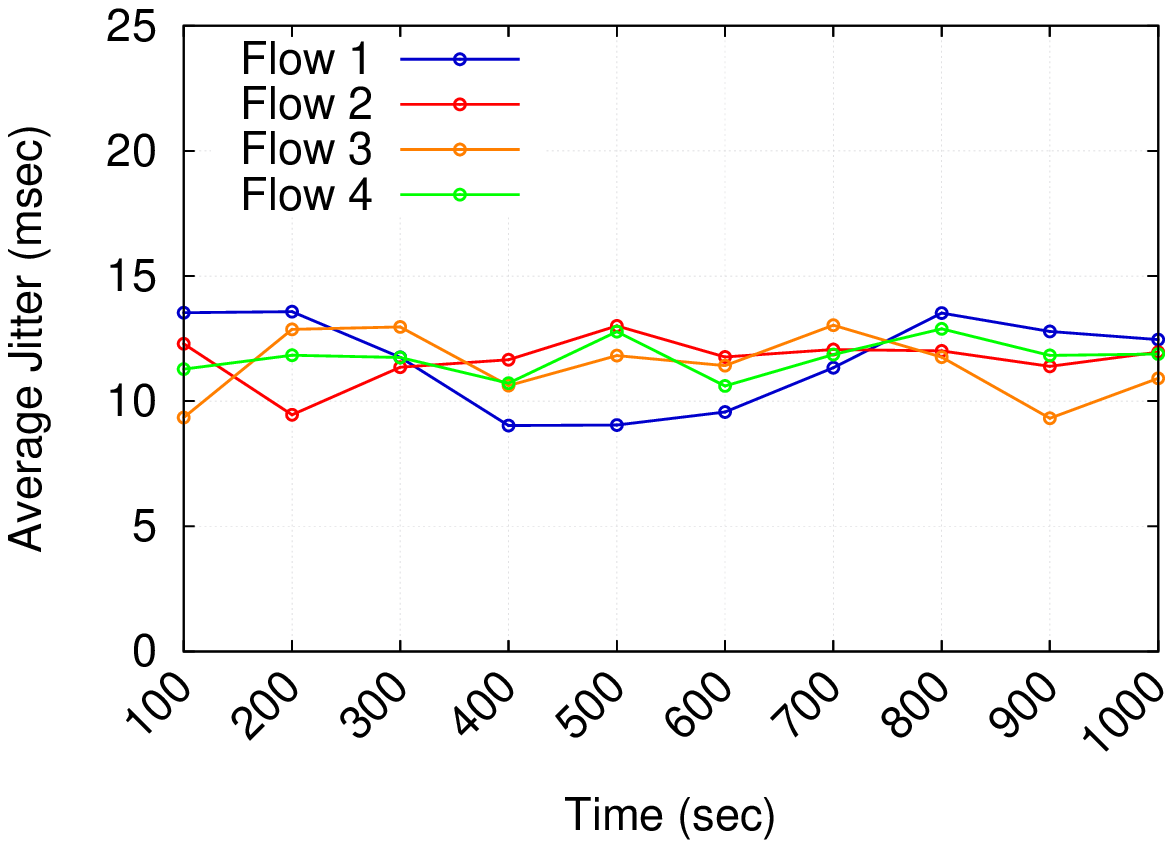}
%\vspace{-0.3cm}
\subcaption{App. unaware}
\label{fig:unaware_delay_2_TCP_UDP}
\end{minipage}
%\vspace{-0.2cm}
\caption {Class priority versus per class jitter, two way UDP and TCP}
%\vspace{-0.5cm}
\end{figure}

% \begin{figure}[htb!]
% \centering
% %\vspace{-0.2cm}
% \epsfig{width=7cm, figure=images/tcp_udp_2_aware_jitter.eps}
% %\vspace{-0.3cm}
% \caption{Application aware: Class priority versus per class jitter, two-way UDP and TCP}
% \label{fig:delay_2_TCP_UDP}
% %\vspace{-0.3cm}
% \end{figure}
This experiment of evaluation uses a similar setup as the previous experiment (4 TCP and 4 UDP flows from H1 to H2 and H2 to H1). Here, jitter is calculated only for UDP flows. The awareness of application exhibits much clearer benefit when the network load is high. Because the jitter and delay tolerance is low only for Class 1, Class 1 experiences significantly lower jitter than the other groups when the network is application-aware as shown in Figure.~\ref{fig:delay_2_TCP_UDP}. On the other hand, fluctuation of jitter was observed to independently of class without application awareness as shown in Figure.~\ref{fig:unaware_delay_2_TCP_UDP}. This makes higher priority class to suffer more jitter.

% \begin{figure}[htb!]
% \centering
% %\vspace{-0.3cm}
% \epsfig{width=7cm, figure=images/tcp_udp_2_unaware_jitter.eps}
% %\vspace{-0.3cm}
% \caption{Application unaware: Class priority versus per class jitter, two way UDP and TCP}
% \label{fig:unaware_delay_2_TCP_UDP}
% %\vspace{-0.3cm}
% \end{figure}

\subsubsection{Throughput analysis using TCP for a new flow started at \textit{t}}
\begin{figure}[htb!]
\centering
%\vspace{-0.3cm}
\epsfig{width=7cm,figure=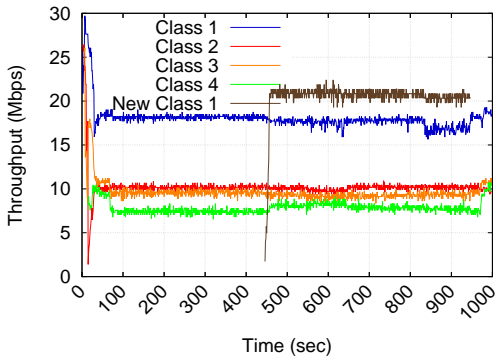}
%\vspace{-0.3cm}
\caption{Application aware: Class priority versus per class TCP throughput, two way TCP with a flow starting at time t=450 sec}
\label{fig:aware_throughput_TCP_newflow}
%\vspace{-0.3cm}
\end{figure}
In Figure.~\ref{fig:aware_throughput_TCP_newflow}, a Class~1 flow is started at \textit{t} = 450 sec. The network is loaded with TCP and UDP flows. For a new flow entering the network the AMPF should find the best path for this Class~1 flow. In Figure.~\ref{fig:aware_throughput_TCP_newflow}, it is evident that AMPF finds the best path for this Class~1 traffic and hence it satisfies the throughput requirement. The throughput of the new flow is above 20 Mbps. This shows intelligence of the network.
%\vspace{-0.3cm}
\section{Acknowledgment}
We thank Bhargav Reddy and Aradhya Biswas for their valuable contribution in this work.
\section{Conclusion and Future Work}

To facilitate the heterogeneous requirement of various applications in enterprise networks, we introduced AMPF using MLT and SDN. AMPF automatically classifies the input traffics and applies QoS policy to each of them. Each traffic travels the most appropriate path to reach the destination to achieve low-latency or higher throughput. Unlike MPLS no new tags are added to the packet in our model. The proof of concept implementation was evaluated in various experiments using mininet emulation. The result exhibited a significant reduction of latency and improvement of throughput to deliver the service for the classified application. It is very clear that unclassified applications did not meet the requirement while the multipath assignment is done. This indicates that we can improve the user experience of various applications that have a variety of demands which network can offer as a QoS service if classified. AMPF will be further extended to large-scale deployment in the campus network at IIT Hyderabad. As part of further work, optimization for flow assigning algorithm can also be done.
% In an enterprise scenario, the traffic is heterogeneous in terms of its delay requirement bandwidth guarantee (QOS). There are many distributed approach to address this requirement. But SDN has the advantage of gathering entire topology of the network; also it has the control over all flows in the network. So, we have chosen SDN approach for improving QOS by sending different traffic through different paths. The multipath transmission of traffic is done in literature, but without having an awareness of the kind of traffic sent. We focused on this particular problem and classified the kind of traffic at the controller and based on the class of traffic multiple paths are assigned to different traffic. It is evident from the results that sensitive flow is taking a better path than insensitive flow. A significant reduction in terms of latency for specific classes of application is done. This, in turn, improves the user experience for sensitive application. The Algorithm which is proposed can be optimized. But, the optimization problem can also become NP- Complete. In such scenario, our heuristic method works well. One can make improvement in the heuristic algorithm to achieve near optimal solution. 
%\vspace{-0.4cm}

%\bibliography{AMPF} {} % sigproc.bib is the name of the Bibliography in this case
%\bibliographystyle{abbrv}

% that's all folks
\end{document}